\documentclass[reprint,
superscriptaddress,
showpacs,
 amsmath,amssymb,
 aps,
 pra,
]{revtex4-2}

\usepackage[english]{babel}
\usepackage{graphicx}
\usepackage{dcolumn}
\usepackage{bm}
\usepackage[pdfstartview=FitH,colorlinks=true,citecolor=blue,linkcolor=blue,urlcolor=blue]{hyperref}
\usepackage[mathlines]{lineno}
\usepackage{xcolor}
\usepackage{tikz}
\usetikzlibrary{patterns}
\usepackage{natbib}
\usepackage{braket}
\usepackage{cleveref}
\usepackage{longtable}
\usepackage{float}
\usepackage{bm}
\usepackage{tikz}
\usetikzlibrary{calc,patterns,decorations.pathmorphing,decorations.markings}
\PassOptionsToPackage{hyphens}{url}

\usepackage{ytableau}
\ytableausetup{textmode, boxframe=0.05em, boxsize=0.5em, centertableaux}

\newcommand{\up}{\uparrow}
\newcommand{\down}{\downarrow}

\newcommand{\hP}{\hat{P}}

\usepackage{cancel}
\usepackage{soul}

\begin{document}

\title{Symmetry oscillations sensitivity to SU(2)-symmetry breaking in quantum mixtures}

\author{S. Musolino}
%\email{silvia.musolino@lpmmc.cnrs.fr}
\affiliation{Universit\'e Grenoble Alpes, CNRS, LPMMC, 38000 Grenoble, France}
\affiliation{Universit\'e C\^ote d'Azur, CNRS, Institut de Physique de Nice, 06200 Nice, France}
%\author{F. Hébert} 
%\affiliation{Universit\'e C\^ote d'Azur, CNRS, Institut de Physique de Nice, 06200 Nice, France}
\author{M. Albert} 
\affiliation{Universit\'e C\^ote d'Azur, CNRS, Institut de Physique de Nice, 06200 Nice, France}
\affiliation{Institut Universitaire de France}
\author{P. Vignolo}
\affiliation{Universit\'e C\^ote d'Azur, CNRS, Institut de Physique de Nice, 06200 Nice, France}
\affiliation{Institut Universitaire de France}
\author{A. Minguzzi}
\affiliation{Universit\'e Grenoble Alpes, CNRS, LPMMC, 38000 Grenoble, France}

\begin{abstract}
In one-dimensional bosonic quantum mixtures with SU(2)-symmetry breaking Hamiltonian, the dynamical evolution explores different particle exchange symmetry sectors.  For the case of infinitely strong intra-species repulsion,  the hallmark of such symmetry oscillations are 
time modulations of the momentum distribution \href{https://doi.org/10.1103/PhysRevLett.133.183402}{[Phys. Rev. Lett. 133, 183402 (2024)]}, an observable routinely accessed in experiments with ultracold atoms.
In this work we show that this phenomenon
is robust in strongly interacting quantum mixtures with arbitrary inter-species to intra-species interaction strength ratio.
Taking as initial state the ground state of the SU(2) symmetric Hamiltonian and  time-evolving with the  symmetry breaking Hamiltonian,
we analyze how the amplitude and frequency of symmetry oscillations, and thus of the momentum distribution oscillations, depend on the strength of the symmetry-breaking perturbation. We find that the set of symmetry sectors which are coupled during the time evolution is dictated by the spin-flip symmetry of the initial state and show that the population of the initial state may
vanish periodically, even in the thermodynamic limit, thus revealing the robustness and universality of the symmetry oscillations.
\end{abstract}

\maketitle

\section{Introduction}\label{sec:intro}
 Permutation symmetries are a cornerstone in the description of many-body quantum systems composed of identical particles. While bosons and spin-polarized fermions are associated with fully symmetric or antisymmetric wave functions, multicomponent systems can exhibit more complex
 symmetry properties under particle exchange. The eigenstates of such systems can be classified according to the irreducible representations of the symmetric group $S_N$, which encapsulates all possible permutations of $N$ particles. Strongly interacting one-dimensional (1D) quantum gases offer a unique platform to probe the role of permutation symmetry, both in static~\cite{art:fang2011, art:decamp2016exact, art:decamp2016_high, art:aupetit2022, art:Harshman_2015, art:Harshman_2016} and dynamical~\cite{art:musolino2024} properties. Their realization with ultracold atoms allows for a high degree of experimental control and tunability, making them ideal for exploring fundamental aspects of quantum statistics and correlations.

In a recent study~\cite{art:musolino2024}, we demonstrated that the momentum distribution of a strongly interacting two-component mixture of hard-core bosons evolves dynamically in a way that reflects oscillations between many-body states of distinct particle-exchange symmetry. These \emph{symmetry oscillations} emerge when the system is initialized in a state that is not an eigenstate of the symmetry-breaking
Hamiltonian, an effect reminiscent of neutrino flavor oscillations~\cite{10.1093/ptep/ptac097}.

In this work, we 
generalize our previous analysis to a broader class of Hamiltonians and initial conditions, in order to unveil the universal features of symmetry oscillations and to identify regimes where the effect becomes particularly pronounced. Specifically, we consider a model with 
different inter- and intra-species
contact
interactions characterized by a symmetry-breaking parameter $\lambda = \frac{g_{\uparrow\downarrow}}{g} - 1,$ where $g_{\uparrow\downarrow}$ and  $g_{\uparrow\uparrow} = g_{\downarrow\downarrow}=g$  are the %inter-species and (equal) intra-species 
interaction strengths.
%, respectively.
For $\lambda = 0$, the system possesses SU$(2)$ symmetry, 
and no symmetry oscillations occur.

When $\lambda \neq 0$, symmetry oscillations  take place,  due to the coupling among  different symmetry sectors.  We show that for small values of $|\lambda|$, a second-order perturbative expansion accurately captures the dynamics, allowing for an analytical understanding of both the frequency and amplitude of the oscillations. This weak symmetry-breaking regime is experimentally relevant in setups where SU$(2)$ symmetry is approximately preserved~\cite{arX:Beugnon2024, arX:Zhao2025}. In contrast, for strong symmetry breaking ($|\lambda| \gg 1$), we observe a complete depletion of the initial symmetry sector and a dynamical evolution that reveals universal features independent of the microscopic details of the symmetry-breaking perturbation.

The paper is organized as follows. In Sec.~\ref{sec:model}, we present the model Hamiltonian and its mapping to an effective spin-chain model. We also introduce relevant symmetry considerations, selection rules, and observables. In Sec.~\ref{sec:res}, we present numerical results for few-body systems as a function of $\lambda$, and distinguish between the perturbative regime (Sec.~\ref{sec:perturbative}) and the strongly symmetry-broken regime (Sec.~\ref{subsec:largel}). In Sec.~\ref{sec:initialcond}, we analyze the effect of different initial states. Finally, Sec.~\ref{sec:conclu} summarizes our main findings and discusses experimental perspectives and further theoretical developments. In Appendix~\ref{app}, we provide explicit expressions for the equations used in Sec.~\ref{sec:perturbative}.

\section{Model}\label{sec:model}
The Hamiltonian for a one-dimensional mixture of $N$ bosonic particles with two equally populated components ($\uparrow$ and $\downarrow$), each of mass $m$, interacting via repulsive contact interactions is given by
\begin{equation}
\begin{split}
H&=\sum_{\sigma=\uparrow,\downarrow} \sum_{i=1}^{N_\sigma} 
\left[-\frac{\hbar^2}{2m}\frac{\partial^2}{\partial x_{i,\sigma}^2} + V(x_{i,\sigma})\right] \\ &+g_{\uparrow\downarrow}\sum_{i, j}\delta(x_{i,\uparrow}-x_{j,\downarrow})  
+ \sum_{\sigma= \uparrow, \downarrow} g_{\sigma\sigma}\sum_{i<j} \delta(x_{i,\sigma}-x_{j,\sigma}),
\end{split}
 \label{eq:ham}
\end{equation}
where $V(x)$ is a spin-independent external trapping potential taken as a hard-wall box trap in our case, and $g_{\sigma\sigma'}$ denotes the inter- ($\sigma \neq \sigma'$) or intra-species ($\sigma = \sigma'$) interaction strength. In the limit $g_{\sigma\sigma'}\rightarrow +\infty$ for any $\sigma,\sigma'$, the many-body wave function $\Psi$ vanishes whenever {$x_{i,\sigma} = x_{j,\sigma'}$}. 

The many-body problem in the strongly repulsive regime can be solved using a generalized time-dependent Bose-Fermi mapping~\cite{art:girardeau-timedependent,art:deuretzbacher2008_lett, art:volosniev_nat,capuzzi2024}.
%Pat: 
For the specific case where the total density is not excited, this mapping
allows the exact time-evolving many-body wave function to be written as
\begin{equation}
\Psi(\vec{X}, \vec{\sigma}, t)= \sum_{P\in S_N}   \braket{\vec{\sigma} |\hat{P}|\chi(t)} \theta_P(\vec{X})\Psi_A(\vec{X}),
\label{eq:Psi_MB}
\end{equation}
where the system is described in terms of spatial ($\vec{X} \equiv \{ x_{1},\dots,x_{N} \}$) and spin ($\vec{\sigma} \equiv \{ \sigma_{1},\dots,\sigma_{N} \}$) degrees of freedom, and where we have identified $\{x_i,\sigma_i\} = x_{i,\sigma}$ as the spatial coordinate and spin label of the $i$-th particle. 
In Eq.~\eqref{eq:Psi_MB}, the operator $\hat P$ denotes a permutation in $S_N$, $\ket{\chi(t)}$ is the spin state at time $t$, and $\theta_P(\vec{X})$ is the generalized Heaviside function, equal to 1 in the coordinate sector $x_{P(1)}<\dots<x_{P(N)}$ and 0 otherwise. The orbital wave function is given by $\Psi_A = A \Psi_F$, with $A = \prod_{i<j} \mathrm{sgn}(x_i - x_j)$ and $\Psi_F = (1/\sqrt{N!}) \det [\phi_j(x_k)]$, where $\Psi_F$ is the Slater determinant wave function for $N$ spinless non-interacting fermions constructed from the 
orbitals $\phi_j(x)$,
eigenstates of the single-particle problem 
in the potential $V(x)$.

\subsection{Mapping to an effective spin chain}\label{subs:hamil}

In the limit of strongly repulsive interactions, Eq.~\eqref{eq:ham} can be mapped onto an effective spin Hamiltonian defined in a Hilbert space of dimension $M=N!/(N_\uparrow!N_\downarrow!)$, where the site index corresponds to the particle index~\cite{art:massignan2015, art:Yang_2016, art:barfknecht_2018, art:aupetit2022, art:zhang_wang}, and the number of sites equals the total number of particles. For $g_{\uparrow\uparrow}=g_{\downarrow\downarrow}= g$ and  $\lambda= g_{\uparrow\downarrow}/g -1$, the Hamiltonian can be written as a sum of a SU(2) symmetric part and a symmetry-breaking perturbation according to:
\begin{equation}
\begin{split}
\hat{H}_{\lambda} &\equiv \hat{H}_\mathrm{SU} + \lambda \hat{V}_\mathrm{SB}\\
&= -\sum_{i=1}^{N-1}  \frac{2\alpha_i}{g_{\uparrow \downarrow}} \left( \mathbf{S}_{i} \cdot \mathbf{S} _{i+1}+\frac{3}{4} {\mathbf{I}}\right) \\ 
&-\lambda \sum_{i=1}^{N-1}  \frac{2\alpha_i}{g_{\uparrow \downarrow}} \left(2{S_{i}^{(z)} S_{i+1}^{(z)}}+\frac{1}{2}  {\mathbf{I}}\right),
\label{eq:Hspin}
\end{split}
\end{equation}
where the zero of the energy is fixed at  the Fermi energy, 
{$\mathbf{I}$ is the identity matrix, $\mathbf{S}_i = (S_i^{(x)}, S_{i}^{(y)}, S_{i}^{(z)})$ is the spin operator at site $i$ expressed in terms of its components,} and the couplings $\alpha_i$ are given by
\begin{equation}
\alpha_i = \frac{ N! \hbar^4}{m^2} \int d\vec{X} \theta_\mathrm{id}(\vec{X})\delta(x_i - x_{i+1}) \Big|\frac{\partial \Psi_A}{\partial x_{i}}\Big|^2. 
\label{eq:Ji}
\end{equation}

Note that when $\lambda=0$ ($g = g_{\uparrow\downarrow}$), the effective spin Hamiltonian (\ref{eq:Hspin}) reduces to a SU(2)-symmetric XXX Heisenberg chain, denoted by {$\hat{H}_{SU}$}. In the general case $\lambda\neq 0$ ($g \neq g_{\uparrow\downarrow}$), $\hat{H}_{\lambda}$ corresponds to an XXZ chain, which explicitly breaks the $SU(2)$ symmetry. A notable special case occurs when $\lambda = -1/2$, for which the Hamiltonian becomes equivalent to an XX model, 
as already observed in Ref.~\cite{art:volosniev_2015}.
Adopting standard notation for spin-chain Hamiltonians~\cite{book:franchini2017}, the anisotropy parameter $\Delta = (1 + 2\lambda)$ quantifies the strength of the uniaxial anisotropy along the $z$ direction. For $|\Delta| < 1$ ($-1 < \lambda < 0$), the $x$-$y$ planar terms dominate. Conversely, for $|\Delta| > 1$ ($\lambda < -1$ or $\lambda > 0$), the axial interaction along $z$ becomes dominant. In particular, $\lambda > 0$ ($\lambda < -1$) corresponds to a ferromagnetic (antiferromagnetic) phase along the $z$ axis. In the intermediate regime $-1 < \lambda < 0$, the system exhibits a paramagnetic behavior.

%snippet basis
 We now consider the so-called snippet basis of size $M$~\cite{art:minguzzi2022strongly}, $\{ \ket{P} \}$, made of all the possible spin-ordered states and  calculate the matrix elements of $\hat H_\lambda$ in this basis. To do so, we first rewrite Eq.~\eqref{eq:Hspin} in terms of  spin matrices, which are $\mathbf{S}_i = \boldsymbol{\sigma}_i /2$ with $\boldsymbol{\sigma} = (\sigma^{(x)}, \sigma^{(y)}, \sigma^{(z)})$ are the Pauli matrix and using the relation $\hat P_{i, i+1} = (\boldsymbol{\sigma}_i \cdot \boldsymbol{\sigma}_{i+1} +  {\mathbf{I}})/2$, we have
 \begin{eqnarray}
   \hat{H}_\mathrm{SU} &=& - \sum_{i=1}^{N-1} \frac{\alpha_i}{ g_{\up \down}} \left( \hat P_{i, i+1} +   {\mathbf{I}} \right),   \label{eq:HSU_Pii1}\\
   \hat{V}_\mathrm{SB} &=& - \sum_{i=1}^{N-1} \frac{\alpha_i}{ g_{\up \down}} \left( \sigma_i^{(z)} \sigma_{i+1}^{(z)}+   {\mathbf{I}} \right), 
   \label{eq:VSB_Pii1}
 \end{eqnarray}
 where it is clear that the perturbation is along the axial direction. In the snippet basis, the elements along the diagonal are given by
  \begin{align}
      [g_{\up \down} \hat{H}_\mathrm{SU}]_{PP}  &= - \sum_{i=1}^{N-1} \alpha_i \delta^{(P)}_{\sigma_i, \sigma_{i+1}} - \sum_{i=1}^{N-1}\alpha_i,
       \label{eq:Hsu_diag}\\
      [g_{\up \down} \hat{V}_\mathrm{SB}]_{PP} 
   &= - \sum_{i=1}^{N-1} \alpha_i F^{(P)}_{\sigma_i, \sigma_{i+1}}  - \sum_{i=1}^{N-1}\alpha_i,
    \label{eq:VSB_diag}
  \end{align}
       where the subscript $(P)$ indicates the corresponding snippet and
 \begin{equation}
 F^{(P)}_{\sigma_i, \sigma_{i+1}}=
     \begin{cases}
      1 & \mathrm{if} \,\,\sigma_i = \sigma_{i+1} \\
      -1 & \mathrm{if} \,\,\sigma_i \neq \sigma_{i+1}
     \end{cases},
     \label{eq:Fdef}
 \end{equation}
 because $\sigma^{(z)}_i \ket{\up}_i = +1 \ket{\up}_i$ and $\sigma^{(z)}_i \ket{\down}_i = -1 \ket{\down}_i$. In the snippet basis, the off-diagonal element of  $[g_{\up \down} \hat{V}_\mathrm{SB}]_{PP'} = 0$ are always zero, because $\braket{m' | \sigma^{(z)} | m} = \pm \delta_{m, m'}$ Therefore, the snippets that are not affected by the symmetry breaking perturbation are the one for which $\sum_{i=1}^{N-1} F^{(P)}_{\sigma_i, \sigma_{i+1}} = -(N-1)$ that are the one of the  type $\ket{P} = \ket{\up \down \up \down \cdots \up \down}$ and its spin-flipped state. The  off-diagonal elements are instead given by
 \begin{equation}
  [g_{\up \down} \hat{H}_\mathrm{SU}]_{PP'} =
  \begin{cases}
   -\alpha_i & \mathrm{if} \,\, \exists \hat P_{i, i+1} :  \hat P_{i, i+1} \ket{P} = \ket{P'}  \\
    & \mathrm{for}\,\, i \in [1, N-1] \\
   0 & \mathrm{otherwise}
  \end{cases}.
     \label{eq:Hsu_offd}
 \end{equation}

 For the box trap considered in this work $\alpha_i  = \alpha= \hbar^4\pi^2 N(N+1)(2N+1)/(6m^2L^3)$ for all $i$ \cite{art:aupetit2023}, the diagonal elements are given by
 \begin{eqnarray}
 \left[\hat{H}_\lambda\right]_{PP}  &=& -\frac{\alpha}{g_{\up\down}} \left(2 N_{\sigma\sigma}^{(P)} + N_{\sigma \bar \sigma}^{(P)}  + 2\lambda  N_{\sigma\sigma}^{(P)} \right),
     \label{eq:Hlambda_diag}
 \end{eqnarray}
 where  $N_{\sigma\sigma}^{(P)}$ and $N_{\sigma \bar \sigma}^{(P)}$ are the number of neighboring pairs with equal and opposite spin in the corresponding basis element, $\ket{P}$, respectively. We notice that to obtain Eq.~\eqref{eq:Hlambda_diag}, we have used  the following relation: $N-1 = N_{\sigma\sigma}^{(P)}+ N_{\sigma \bar \sigma}^{(P)}$, which will be useful later. 
 
 We now provide the explicit example for $N=4$, which will be extensively used throughout this work. Using the basis $\{ \ket{\up \up \down \down}, \ket{\up \down \up \down}, \ket{\up \down \down \up}, \ket{\down \up \up \down}, \ket{\down \up \down \up}, \ket{\down \down \up \up}\}$, we obtain
\begin{equation}
\left[\hat{H}_\lambda\right] = - \frac{\alpha}{g_{\up \down}} 
\begin{pmatrix}
5+4\lambda & 1 & 0 & 0 & 0 & 0 \\
1 & 3 & 1 & 1 & 0 & 0 \\
0 & 1 & 4+2\lambda & 0 & 1 & 0 \\
0 & 1 & 0 & 4+2\lambda & 1 & 0 \\
0 & 0 & 1 & 1 & 3 & 1 \\
0 & 0 & 0 & 0 & 1 & 5+4\lambda
\end{pmatrix}.
\label{eq:Hz_pert}
\end{equation}

\subsection{Symmetry and selection rules}\label{subs:sym}

In order to characterize the symmetry of the mixture,  
we use a set of generators of the $SU(2)$ algebra associated with 
the permutation symmetry group. In particular, we use as a symmetry witness the two-cycle class-sum operator, which corresponds to the group partition 
associated with the sum over all transpositions $\hat{P}_{i,j}$~\cite{art:fang2011, art:decamp2016_high, art:aupetit2022}, namely
\begin{equation}
\hat{\Gamma}^{(2)} = \sum_{i<j} \hat{P}_{i, j}.
\label{eq:Gamma2}
\end{equation} 
The eigenstates of $\hat\Gamma^{(2)}$, which can be chosen to be simultaneous eigenstates of ${\hat H}_ {\rm SU}$ since $[\hat H_\mathrm{SU}, \hat{\Gamma}^{(2)}]=0$, denoted $\{\ket{\chi_\ell^{\rm SU}}\}$, can be classified according to Young tableaux, reflecting nontrivial permutation symmetries. For the bosonic systems considered here, the Young tableaux take the form $(N-q, q)$ with $0\le q\le N/2$, where $N-q$ is the number of boxes in the first row and $q$ in the second row. 
The  $\hat\Gamma^{(2)}$-eigenvalue of these states reads
\begin{equation}
\begin{split}
    \gamma_\ell&=\dfrac{1}{2}(N^2-N+2q^2-2q(N+1))\\
    &=\gamma_0-q(N-q+1),
    \label{eq:eigengamma}
    \end{split}
\end{equation}
where we have used the property that connects the number of boxes $\mu_i$ per line $i$ to the eigenvalues of $\hat\Gamma^{(2)}$, $\gamma_\ell=\frac{1}{2}\sum_i\mu_i(\mu_i-2i+1)$ \cite{art:Katriel_1993,art:decamp2016exact},
and defined $\gamma_0=\frac{1}{2}(N^ 2-N)$ as the eigenvalue of the most symmetric state with $q=0$.

In contrast, the eigenstates of $\hat H_\lambda$ with $\lambda\ne 0$, $\{ \ket{\chi_\ell^{(\lambda)}}\}$, do not exhibit a well-defined symmetry but can be expressed as linear combinations of the common  eigenstates of $\hat H_{\rm SU}$ and $\hat\Gamma^{(2)}$.
This symmetry coupling underlies the phenomenon of symmetry oscillations during the dynamical evolution of the quantum mixture governed by $\hat{H}_\lambda$~\cite{art:musolino2024}.

Here, we show that for a balanced mixture ($N_\uparrow=N_\downarrow$) with $g_{\uparrow\uparrow}=g_{\downarrow\downarrow}$, the symmetry coupling induced by the SB term obeys a specific selection rule. Indeed, $\hat H_\lambda$ is invariant under total spin flip, $\hat U = \prod_{i=1}^N \boldsymbol{\sigma}_i^{(x)}$, for any $\lambda$, i.e., $\hat U^\dagger \hat H_{\lambda} \hat U=\hat H_{\lambda}$. 
This allows one to choose a basis that diagonalizes both $\hat H_{\lambda}$ and $\hat U$ simultaneously, such that $\hat U |\chi_n^{(\lambda)}\rangle=\pm |\chi_n^{(\lambda)}\rangle$, with $\pm 1$ the eigenvalues of the unitary operator $\hat U$.
In particular, we note that the eigenstates of $\hat H_{\rm SU}$ satisfy the following property:
\begin{equation}
    \hat U \ket{\chi^{(N-q,q)}}=(-1)^q \ket{\chi^{(N-q,q)}},
    \label{eq:Uchi}
\end{equation}
where $\ket{\chi^{(N-q,q)}}$ denotes a given eigenstate $\ket{\chi^{\rm SU}_\ell}$ with symmetry $(N-q, q)$, and $q$ is the number of antisymmetric exchanges. 
This implies that the eigenstates $|\chi_\ell^{(\lambda)}\rangle$ are linear combinations of states associated with Young tableaux having either even or odd $q$, but not both. This result is consistent with Ref.~\cite{art:Yurovsky2014}, where the authors show that for spin-$1/2$ systems with $q$-body spin- or position-dependent interactions, the coupled diagrams differ by $q$ boxes being moved from one row to another.
 
\subsection{Physical observables and symmetry}\label{subs:obs}
By following Ref.~\cite{art:musolino2024}, we introduce a momentum density operator, $\hat n_k$, such that the total momentum distribution is given by
$n(k, t)=  \bra{\chi(t)} \hat n_k  \ket{\chi(t)}$ and  is defined as follows:
\begin{equation}
\hat n_k = \sum_{i, j=1}^N  \hat{P}_{(i,..,j)}   R^{(i, j)}(k), 
\label{eq:nk}
\end{equation}
where $\hat{P}_{(i,...,j)}$ is the cyclic (anticyclic)  permutation $i \to i+1 \to \cdots \to j-1 \to j \to i$ ($i \to i-1 \to \cdots \to j+1 \to j \to i$) if $i<j$ ($i>j$) and the identity if $i=j$. The orbital part of Eq.~\eqref{eq:nk} is given by
\begin{equation}
\begin{split}
R^{(i, j)}(k) &= \frac{N!}{2\pi}\int dx dx'  e^{-ik(x-x')} \displaystyle\int_{I_{ij}} \left(\prod_{n\neq i} dx_n \right)\\
& \Psi_\mathrm{A}(x_1, \dots, x_{i-1}, x, x_{i+1}, \cdots, x_N)\\ &\times  \Psi_\mathrm{A}(x_1, \dots, x_{i-1}, x', x_{i+1}, \cdots, x_N),
\end{split}
\label{eq:rhoij}
\end{equation} 
with $I_{ij}$ is the integration interval defined by $x_1 < \cdots < x_{i-1} < x < x_{i+1} < \cdots < x_{j-1} < x' < x_{j} < \cdots < x_N$ ($i<j$)~\cite{art:deuretzbacher2016_num}. The crucial result of  Ref.~\cite{art:musolino2024} was to proof that
\begin{equation}
[\hat{\Gamma}^{(2)},\hat n _k ]  = 0,
    \label{eq:comm_HJ}
\end{equation}
namely at every $k$ the momentum distribution can be decomposed in terms of symmetry components, as follows:
\begin{equation}
n(k, t) = \sum_{\ell=1}^{M} |\braket{\chi(t)| \xi_\ell(k)}|^2 n_\ell(k),
\label{eq:nk_decomp}
\end{equation}
 where the basis $\{\ket{\xi_\ell(k)}\}$ diagonalizes both $\hat n_k$ and $\hat \Gamma^{(2)}$, such that $\hat n_k  \ket{\xi_\ell(k)} = n_\ell(k) \ket{\xi_\ell(k)}$ and $\hat{\Gamma}^{(2)} \ket{\xi_\ell(k)} = \gamma_\ell \ket{\xi_\ell(k)}$.

We now consider the time evolution of the spin state $\ket{\chi(t)}$ 
\begin{equation}
    \ket{\chi(t)} = e^{-i \hat{H}_\lambda t/\hbar} \ket{\chi(0)},
    \label{eq:time_evolution}
\end{equation}
for which the expectation value of $\hat{\Gamma}^{(2)}$, i.e., $\gamma^{(2)}(t) = \braket{\chi(t) |\hat{\Gamma}^{(2)}| \chi(t)}$, oscillates according to the following equation:
\begin{equation}
\begin{split}
\gamma^{(2)}(t) &= \sum_{m, n} s_m(\lambda) s_n(\lambda) \sum_{\ell} \braket{\chi_n^{(\lambda)}|\chi_\ell^\mathrm{SU} }\gamma_\ell \\&\times \braket{ \chi_\ell^\mathrm{SU} | \chi_m^{(\lambda)}}e^{-i\nu_{mn}(\lambda)t}\\
&=\sum_{\ell}\gamma_\ell |w_\ell (t)|^2,
\end{split}
\label{eq:gamma2_exp}
\end{equation}
with $s_n(\lambda) = \braket{\chi_n^{(\lambda)}|\chi(0)} \in \mathbb{R}$, $\nu_{mn}(\lambda) = (\epsilon_m(\lambda)-\epsilon_n(\lambda))/\hbar$  with $\epsilon_n(\lambda)$ the eigenvalues of $\hat H_\lambda$,  and the symmetry amplitudes are defined as
\begin{equation}
    w_\ell (t) \equiv \sum_n s_n(\lambda) \braket{\chi_\ell^\mathrm{SU} | \chi_n^{(\lambda)}} e^{-i \epsilon_n(\lambda) t/\hbar}.
    \label{eq:w_l}
\end{equation}
The symmetry state population of a given symmetry $\gamma$ is given by ${\cal W}_\gamma(t)=\sum_{\ell \in \gamma} |w_\ell(t)|^2$. 
Using the $\ket{\chi_\ell^{\rm SU}}$ basis, the momentum distribution (Eq.~\eqref{eq:nk_decomp}) can be written
\begin{equation}
n(k, t) = \sum_{\ell, \ell'} \braket{\chi_\ell^{\rm SU} |\hat n_k | \chi_{\ell'}^{\rm SU} } w_\ell^*(t) w_{\ell'}(t),
\label{eq:nk_decomp_wl}
\end{equation}
showing that, with respect to the decomposition in Eq.~\eqref{eq:nk_decomp}, is not diagonal for a generic $k$. However, as pointed out in Ref.~\cite{art:musolino2024}, in the limit of large $k$, Eq.~\eqref{eq:nk_decomp_wl} becomes diagonal. In such a limit, $\lim_{k \to \infty} k^4 n(k, t)= \mathcal{C}(t)$, where $\mathcal{C}(t)$ is dominated by the Tan contact, $\mathcal{C}_T(t)$ with eventual additional size  effects in case of nonsmooth trapping potentials~\cite{art:aupetit2023, art:musolino2025}. Indeed, the large-$k$ tails of the momentum distribution can be connected to the symmetry state populations 
 (see Supplemental materials of Ref.~\cite{art:musolino2024}):
\begin{equation}
\begin{split}
  \mathcal{C}_T(t) &= - \frac{m^2}{\pi \hbar^4}
  \Big(\braket{ \chi(t) |\frac{\partial \hat{H}_\lambda}{\partial 1/g_{\up\down}} | \chi(t)}+\braket{ \chi(t) |\frac{\partial \hat{H}_\lambda}{\partial 1/g} | \chi(t)}\Big) \\
  &=
  \frac{m^2}{\pi \hbar^4} \sum_{i=1}^{N-1} 2\alpha_i  \bra{\chi(t)} \Big({{\bf S}_{i}\cdot {\bf S} _{i+1}}  +\frac{3}{4}   {\mathbf{I}} \Big)\ket{\chi(t)}\\
  &= \frac{m^2}{\pi \hbar^4} \sum_{i=1}^{N-1} \alpha_i  \braket{\chi(t)| \left(\hP_{i, i+1} +    {\mathbf{I}} \right)| \chi(t)},
   \end{split}
   \label{eq:adiab_time}
\end{equation}
where the last line reads as the expectation value of the SU Hamiltonian (Eq.~\eqref{eq:HSU_Pii1}) in the eigenvectors of the total Hamiltonian $\hat H_\lambda$. Indeed, one can write 
\begin{equation}
    \begin{split}
        \mathcal{C}_T(t) &=  \sum_{m, n} s^\ast_m(\lambda) s_n(\lambda) e^{i \nu_{mn}(\lambda) t} \\&\times \sum_\ell \braket{\chi_m^{(\lambda)} |\chi_\ell^\mathrm{SU}} c_\ell^\mathrm{SU} \braket{ \chi_\ell^\mathrm{SU} |\chi_n^{(\lambda)}}\\
        &= \sum_{\ell} c_\ell^{\rm SU} |w_\ell(t)|^2,
    \end{split}
    \label{eq:CT_decomp}
\end{equation}
where $c_\ell^\mathrm{SU} =  -m^2 g_{\up \down}  \epsilon_\ell^\mathrm{SU}/(\pi \hbar^4)$ depends on the eigenvalues $\epsilon_\ell^\mathrm{SU}$ of 
$ \hat{H}_\mathrm{SU}$.

\section{Results}\label{sec:res}

 In Ref.~\cite{art:musolino2024}, we studied the spin dynamics of bosonic mixtures prepared in an initial domain-wall state and subject to a Hamiltonian with $\lambda = -1$ (corresponding to a large, finite $g_{\up\down}$ and $1/g = 0$). In this Section, we extend the analysis to arbitrary values of $\lambda$ and consider initial states with well-defined symmetry. Specifically, we focus on a scenario in which the initial state is the ground state of the SU(2)-symmetric Hamiltonian, that is, $\ket{\chi(0)} = \ket{\chi_0^\mathrm{SU}}$, related to the most symmetric eigenvalues of the class-sum operator $\hat{\Gamma}^{(2)}$, $\gamma_0$.

\begin{figure}
    \centering
     \includegraphics[scale=0.6]{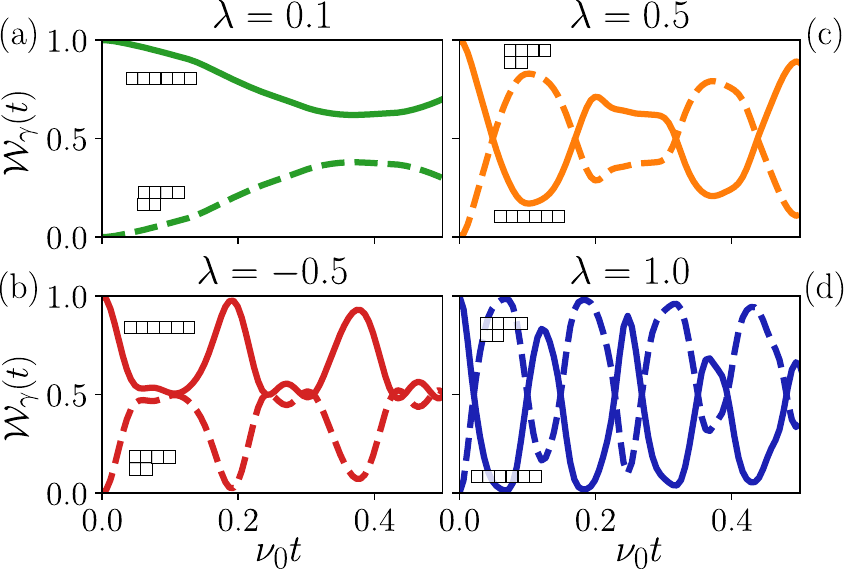}
\caption{Time evolution of the symmetry population ${\cal W}_\gamma(t)$ for the symmetry  $(6) = \ydiagram{6}$ corresponding to the nondegenerate state $\ket{\chi^{\rm SU}_0}$ (solid lines) and  
$(4, 2) = \ydiagram{4,2}$  corresponding to $9$-fold degenerate states of  $\ket{\chi^{\rm SU}_n}$  (dashed lines) for $N=6$ and  $\lambda = 0.1$ (a), $-0.5$ (b), $0.5$ (c), and $1.0$ (d). The other symmetry state populations, related to $\ydiagram{5, 1}$ and $\ydiagram{3, 3}$, which are not shown, remain zero for all times. The
time is in units of $\nu_0 =\hbar/(m L^2 )$ and the interaction strength $g=20(\hbar^2 /m L)$.}
\label{fig:lambda_N6}
\end{figure}

 Figure~\ref{fig:lambda_N6} shows the symmetry state populations ${ \cal W}_\gamma(t)$, which enter the expression for the expectation value of $\hat\Gamma^{(2)}$ (see Eq.~\eqref{eq:gamma2_exp})
 for a system of $N=6$ bosons and various values of $\lambda$. For $N=6$, the most symmetric state corresponds to the Young tableau $\ydiagram{6}$, whose population begins to decrease immediately once the system starts evolving, as a consequence of the symmetry breaking. Simultaneously, the population associated with the symmetry $\ydiagram{4, 2}$ begins to grow. This is the only symmetry that can be coupled to the chosen initial state, due to the selection rules discussed in Sec.~\ref{subs:sym}. Indeed, the other allowed symmetries for six bosons, namely $\ydiagram{5, 1}$ and $\ydiagram{3, 3}$, remain unpopulated, as verified numerically. During the time evolution, the populations of the $\ydiagram{6}$ and $\ydiagram{4,2}$ symmetry sectors exhibit oscillations, with an amplitude that, at small $|\lambda|$ increases with $|\lambda|$, culminating in a full periodic swapping at $\lambda=1$ [see Fig.~\ref{fig:lambda_N6}(d)].

\begin{figure}
    \centering
    \includegraphics[scale=0.6]{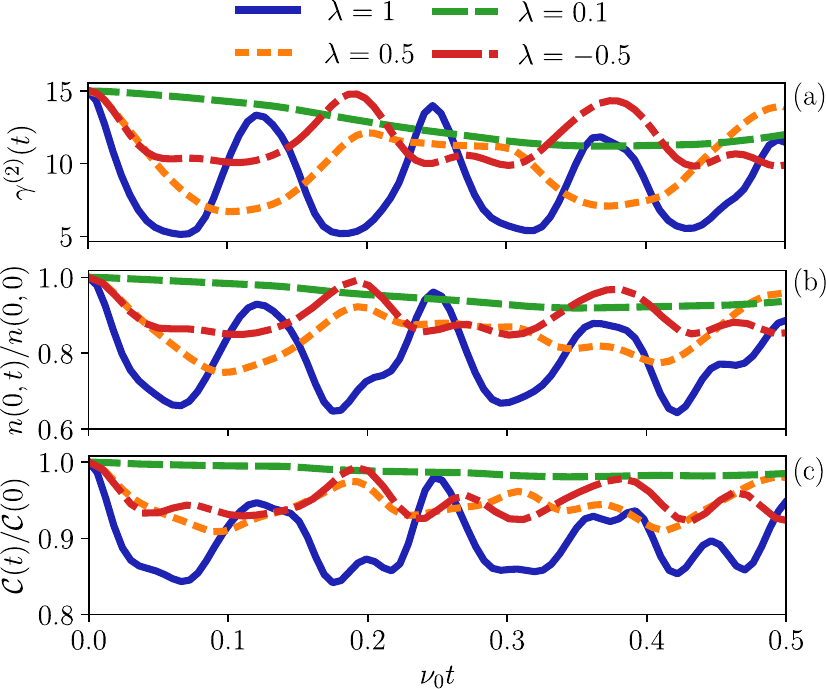}
    \caption{Time evolution of (a) the expectation value of the symmetry witness,  $\gamma^{(2)}(t)$, (b) the peak at $k=0$ of the momentum distribution, $n(0, t)$, and (c) the weight of large momentum tails of the momentum distribution, $\mathcal{C}(t) = \lim_{k\to \infty} k^4 n(k, t)$,  for $N=6$ bosons and different $\lambda$, as indicated in the legend. For all the cases, $\mathcal{C}(t)$ has been obtained numerically by fitting $k^4 n(k, t)$ in the range $kL =60-90$. }
    \label{fig:observables}
\end{figure}

We then explore the dynamical behavior of the symmetry witness (Eq.~\eqref{eq:Gamma2}) and the momentum distribution (Eq.~\eqref{eq:nk}) for different values of $\lambda$. Similarly to Ref.~\cite{art:musolino2024}, Fig.~\ref{fig:observables} shows how the expectation value of the symmetry witness oscillates in time due to the non-conserved symmetry  and these oscillations clearly appear in both the zero-momentum and large-momentum behavior of the momentum distribution, which are experimentally observables.  Remark that for $\lambda=1$, the symmetry witness oscillates between 15 and 5, that is the maximum symmetry oscillation amplitude that can be expected, 15 being the eigenvalue of $\hat\Gamma^ {(2)}$ related to the symmetry $\ydiagram{6}$ and 5 the one related to the symmetry $\ydiagram{4,2}$ (see Eq.~(\ref{eq:eigengamma})).
In such a case, the relative amplitudes of the oscillation of the peak and of the tails of the momentum distribution are $\sim 35\%$ and $\sim 15\%$, respectively.   

In the following sections, we separate the analysis into two regimes: small values of $\lambda$, where the symmetry-breaking term $\lambda \hat V_{\rm SB}$ can be treated as a perturbation of the SU(2)-symmetric Hamiltonian $\hat H_{\rm SU}$ (Sec.~\ref{sec:perturbative}); and large values of $\lambda$, where the symmetry-breaking term becomes dominant (Sec.~\ref{subsec:largel}).

\subsection{Small-$\lambda$ regime}\label{sec:perturbative} 

In the regime $|\lambda| \ll 1$, perturbation theory can be employed to derive analytical expressions for the eigenvalues of  $\hat H_{\rm SU}$. To second order, the eigenvalues of $\hat H_{\rm SU}$ are given by~\cite{book:sakurai}
\begin{equation}
\begin{split}
\epsilon_n(\lambda)& = \epsilon_n^\mathrm{SU} +\lambda \braket{\chi_n^\mathrm{SU} | \hat{V}_\mathrm{SB} | \chi_n^\mathrm{SU} }\\ 
&+ \lambda^2 \sum_{m\neq n} \frac{|\braket{\chi_m^\mathrm{SU} | \hat{V}_\mathrm{SB} | \chi_n^\mathrm{SU}}|^2}{\epsilon_n^\mathrm{SU}  - \epsilon_m^\mathrm{SU} }+ o (\lambda^3),
\end{split}
\label{eq:eigenvalues}
\end{equation}
and the corresponding normalized eigenvectors are given by
\begin{equation}
\begin{split}
\ket{\chi_n^{(\lambda)}} &= \Big(1 - \frac{\lambda^2}{2} \sum_{m \neq n} \frac{|\braket{\chi_m^\mathrm{SU} | \hat{V}_\mathrm{SB} | \chi_n^\mathrm{SU} }|^2}{(\epsilon_n^\mathrm{SU}- \epsilon_m^\mathrm{SU})^2}\Big) \ket{\chi_n^\mathrm{SU}} \\
&+ \lambda \sum_{m \neq n} \frac{\braket{\chi_m^\mathrm{SU} | \hat{V}_\mathrm{SB} | \chi_n^\mathrm{SU} }}{\epsilon_n^\mathrm{SU} - \epsilon_m^\mathrm{SU}} \ket{\chi_m^\mathrm{SU}} \\
&+ \lambda^2 \Bigg( \sum_{\substack{m \neq n \\ l \neq n} } \frac{\braket{\chi_m^\mathrm{SU} | \hat{V}_\mathrm{SB} | \chi_l^\mathrm{SU} } \braket{\chi_l^\mathrm{SU} | \hat{V}_\mathrm{SB} | \chi_n^\mathrm{SU} }}{(\epsilon_n^\mathrm{SU} - \epsilon_m^\mathrm{SU})(\epsilon_n^\mathrm{SU} - \epsilon_l^\mathrm{SU})} \ket{\chi_m^\mathrm{SU}} \\
&\quad - \sum_{m \neq n} \frac{\braket{\chi_m^\mathrm{SU} | \hat{V}_\mathrm{SB} | \chi_n^\mathrm{SU} } \braket{\chi_n^\mathrm{SU} | \hat{V}_\mathrm{SB} | \chi_n^\mathrm{SU} }}{(\epsilon_n^\mathrm{SU} - \epsilon_m^\mathrm{SU})^2} \ket{\chi_m^\mathrm{SU}} \Bigg) \\
&+ o(\lambda^3).
\end{split}
\label{eq:chinlam}
\end{equation}
For $N=4$, we use the SU(2)-eigenvector basis provided in Appendix~\ref{app}, with the corresponding eigenvalues $\epsilon_n^\mathrm{SU} =  (- \alpha/ g_{\uparrow\downarrow})\{6, 4+\sqrt{2}, 3+\sqrt{3}, 4, 4-\sqrt{2}, 3-\sqrt{3}\}$ and associated $\hat\Gamma^{(2)}$ eigenvalues $\gamma_n = \{6, 2, 0, 2, 2, 0\}$ corresponding to  $\{ \ydiagram{4}, \ydiagram{3,1}, \ydiagram{2,2}, \ydiagram{3,1}, \ydiagram{3,1}, \ydiagram{2,2} \}$. 

In a 2+2 bosonic mixture, the highest $\hat\Gamma^{(2)}$ eigenvalue, which is non-degenerate, corresponds to the fully symmetric state, $\gamma_0 = 6 $ ($ \ydiagram{4}$). The intermediate eigenvalue $\gamma_1 = \gamma_3 = \gamma_4 = 2$ ($ \ydiagram{3,1}$) is threefold degenerate and represents states of mixed symmetry. The lowest eigenvalue $\gamma_2 = \gamma_5 = 0 $ ($\ydiagram{2,2}$), which is twofold degenerate, corresponds to the most antisymmetric states.

\begin{figure}
    \centering
    \includegraphics[scale=.6]{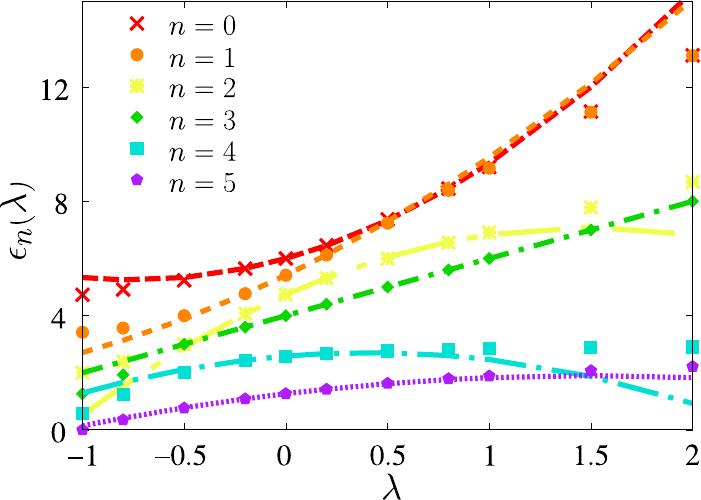}
\caption{Modulus of the eigenvalues of the $H_\lambda$ Hamiltonian  in units of $g_{\up\down}/\alpha$ as a function of $\lambda$ for $N=4$ (Eq.~\eqref{eq:Hz_pert}). Comparison between exact  (dots) and second-order perturbative (lines) results using Eqs.~\eqref{eq:eigenvalues}.}
\label{fig:eigenvalues_N4}
\end{figure}

Figure~\ref{fig:eigenvalues_N4} shows that the perturbative results obtained using Eq.~(\ref{eq:eigenvalues}) (see also Eqs.~\eqref{eq:eigenvalues_N4} in Appendix~\ref{app}) remain accurate well beyond the small-$\lambda$ regime. We also observe that both the perturbative and exact energy levels cluster into three groups at larger $\lambda$, as will be discussed in the next section.

If we consider a fully symmetric initial state, namely $\ket{\chi(0)} = \ket{\chi_0^\mathrm{SU}}$, 
the coefficients $s_n(\lambda)$, obtained by considering the second order expansion of $|\chi_n^{(\lambda)}\rangle$ (see Eqs.~(\ref{eq:eigenvec_2nd_norm})) are given by
\begin{equation}
s_n(\lambda) =
\begin{pmatrix}
1 - 0.44 \lambda^2\\
0 \\
-0.91\lambda -0.83 \lambda^2\\
0\\
0\\
-0.24 \lambda +0.06 \lambda^2 
\end{pmatrix}
+ o (\lambda^3),
\label{eq:sn_lambda}
\end{equation} 
and the only nonzero amplitudes $w_\ell (t)$ (see Eq.~\eqref{eq:w_l}) are $w_{0}(t)$, $w_{2}(t)$, and $w_{5}(t)$ corresponding to the amplitudes of the  $\ydiagram{4}$ and $\ydiagram{2, 2}$ symmetry. To second order in perturbation theory, the latter are given by
\begin{equation}
    \begin{split}
        w_0(t) & = s_0(\lambda)^2 e^{-i \epsilon_0(\lambda) t/\hbar} + s_2(\lambda)^2 e^{-i \epsilon_2(\lambda) t/\hbar} \\&+ s_5(\lambda)^2 e^{-i \epsilon_5(\lambda) t/\hbar},\\
        w_2(t) & = -s_0(\lambda)s_2(\lambda) e^{-i \epsilon_0(\lambda) t/\hbar} + (1-0.41 \lambda^2)s_2(\lambda)  \\&\times e^{-i \epsilon_2(\lambda) t/\hbar} + 0.08 \lambda^2 s_5(\lambda)e^{-i \epsilon_5(\lambda) t/\hbar},\\
         w_5(t) & = -s_0(\lambda)s_5(\lambda) e^{-i \epsilon_0(\lambda) t/\hbar} -0.30 \lambda^2 s_2(\lambda)  \\&\times e^{-i \epsilon_2(\lambda) t/\hbar} + (1-0.03 \lambda^2) s_5(\lambda)e^{-i \epsilon_5(\lambda) t/\hbar}.\\
    \end{split}
    \label{eq:wei_pert_N4}
\end{equation}
Therefore, using Eq.~\eqref{eq:wei_pert_N4}, the expectation value of the symmetry witness (Eq.~\eqref{eq:gamma2_exp}) 
 becomes
\begin{equation}
\begin{split}
\gamma^{(2)}(t)&=\gamma_0|w_0(t)|^2 \\ &=  \gamma_0 (s_0(\lambda)^4 + s_2(\lambda)^4 +  s_5(\lambda)^4) \\
&+ 2\gamma_0 [s_0(\lambda)^2 s_2(\lambda)^2 \cos(\nu_{02}(\lambda) t) \\
&+ s_0(\lambda)^2 s_5(\lambda)^2 \cos(\nu_{05}(\lambda) t) \\
&+ s_2(\lambda)^2 s_5(\lambda)^2 \cos(\nu_{25}(\lambda) t)] ,
\end{split} 
\label{eq:gamma2_pert}
\end{equation}
where we have used that the $\hat\Gamma^{(2)}$-eigenvalues related to $\ket{\chi_2^{\rm SU}}$ and $\ket{\chi_5^{\rm SU}}$ are zeros and therefore do not contribute to the symmetry oscillations of $\gamma^{(2)}(t)$.  We notice that the frequency differences $\nu_{02}(0) =(\epsilon_2^\mathrm{SU}- \epsilon_0^\mathrm{SU})/\hbar $ and $\nu_{05}(0)= (\epsilon_5^\mathrm{SU}- \epsilon_0^\mathrm{SU})/\hbar$ are related to oscillations between the symmetries $\ydiagram{4}$ and $\ydiagram{2,2}$, and $\nu_{25}(0)$ to oscillations between states of the same symmetry $\ydiagram{2,2}$.

\begin{figure}
    \centering
    \includegraphics[scale=.7]{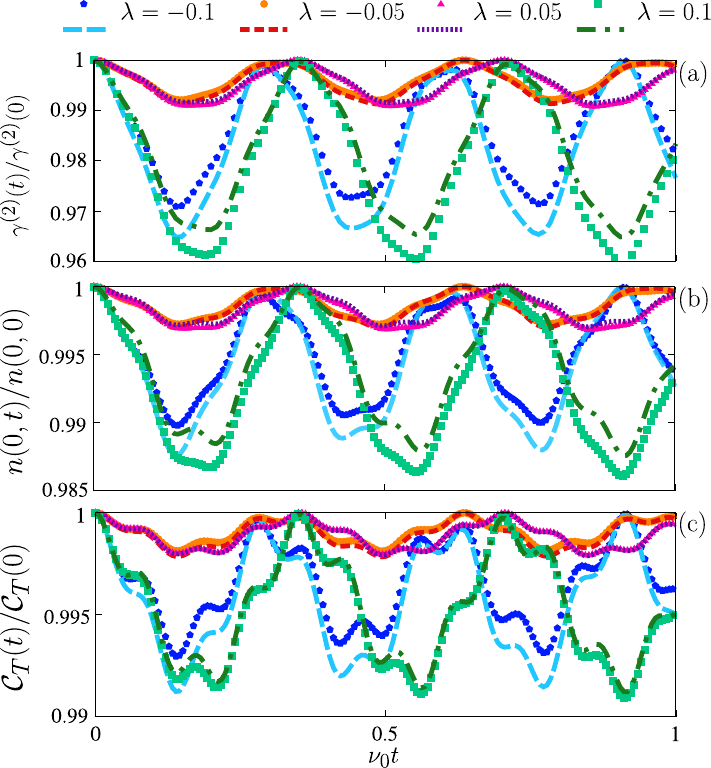}
\caption{Time evolution of (a) the expectation values of $\hat\Gamma^{(2)}$, (b) the peak at $k=0$ of the momentum distribution, $n(0, t)$, and (c) the Tan contact $\mathcal{C}_\mathrm{T}(t)$, normalized by their values at $t=0$, for $N=4$ bosons and small $\lambda$, as indicated in the legend. Comparison between exact  (dots) and second-order perturbative (lines) results using Eqs.~\eqref{eq:gamma2_pert}, \eqref{eq:n0_N4}, and \eqref{eq:CT_N4}. The exact results for the Tan contact have been obtained by fitting $k^4 n(k, t)$ at large $k$ ($kL =60-90$) and subtracting the finite size term~\cite{art:aupetit2023, art:musolino2025}.}
\label{fig:gamma2_N4}
\end{figure}

By performing the perturbative calculation to  second order in $\lambda$, we find that
%Indeed we find that
\begin{equation}
\begin{split}
    \dfrac{\gamma^{(2)}(t)}{\gamma^{(2)}(0)}&\simeq 1+(-1.771+1.656\cos(\nu_{02}(\lambda)t)\\&+0.115\cos(\nu_{05}(\lambda)t))\lambda^2+o(\lambda^3),
    \label{eq:gamma2_N4}
    \end{split}
\end{equation}
namely only the frequencies $\nu_{02}(\lambda)$ and $\nu_{05}(\lambda)$,  which, for $\lambda=0$, represent the energy difference between different symmetry sectors, contributes at this order, with an amplitude that is much larger for the $\nu_{02}(\lambda)$ term.
The relative amplitude of the $\nu_{25}(\lambda)$ frequency, which, for $\lambda=0$, corresponds to an energy oscillation between the same symmetry sector contributes at higher orders of the expansion. As expected, the oscillation amplitude increases as the strength of the perturbation $|\lambda|$ grows. 
This is shown in Fig.~\ref{fig:gamma2_N4}(a), where we compare the time evolution of $\gamma^{(2)}(t)/\gamma^{(2)}(0)$ obtained from the exact solution and from the perturbative expression in Eq.~\eqref{eq:gamma2_N4}. We can observe that already at $|\lambda| = 0.1$, noticeable deviations between the two approaches begin to emerge. 

We can perform a similar analysis for the peak of the momentum distribution $n(0,t)$ and the Tan contact $\mathcal{C}_T(t)$.
$n(0,t)$ is not diagonal in the common eigenbasis of $\hat{H}_\lambda$ and $\hat{\Gamma}^{(2)}$,
but it is still block diagonal, because of the coupling between states belonging to the same symmetry sector \cite{art:musolino2024}.
Indeed, for the case of 2+2 bosons, taking into account the spin-flip symmetry of the chosen initial state, Eq. (\ref{eq:nk_decomp_wl}) simplifies to
\begin{equation}
\begin{split}
    n(0, t) &= \langle\chi_0^{\rm SU}|\hat n_0|\chi_0^{\rm SU}\rangle|w_0(t)|^ 2+ \langle\chi_2^{\rm SU}|\hat n_0|\chi_2^{\rm SU}\rangle|w_2(t)|^ 2\\
   & + \langle\chi_5^{\rm SU}|\hat n_0|\chi_5^{\rm SU}\rangle|w_5(t)|^ 2\\
   &+2\langle\chi_2^{\rm SU}|\hat n_0|\chi_5^{\rm SU}\rangle{\rm Re}[ w_2^*(t)w_5(t)].
    \end{split}
    \label{eq:n0_N4}
\end{equation}
By calculating numerically the terms $\langle\chi_\ell^{\rm SU}|\hat n_0|\chi_{\ell'}^\mathrm{SU}\rangle$ and by using the expansion of the $w_\ell$ amplitudes (see Eqs.\eqref{eq:wei_pert_N4} and Eqs.~\eqref{eq:w0w0}-\eqref{eq:w2w5} of the appendix~\ref{app}), we get
\begin{equation}
\begin{split}
    \dfrac{n(0,t)}{n(0,0)}&=|w_0(t)|^ 2+0.66 |w_2(t)|^ 2+0.45|w_5(t)|^ 2\\&-0.017{\rm Re}[ w_2^*(t) w_5(t)]\\
    &\simeq 1+(-0.592+0.563 \cos(\nu_{02}(\lambda)t)\\&+0.067\cos(\nu_{05}(\lambda)t)-0.038\cos(\nu_{25}(\lambda)t))\lambda^2\\&+o(\lambda^3).
    \label{eq:n0_N4b}
    \end{split}
\end{equation}
Here, due to the coupling between the two states belonging to the same symmetry sector $\ydiagram{2,2}$, the $\nu_{25}(\lambda)$-frequency oscillation contributes at the same perturbative order in $\lambda$ as the $\nu_{02}(\lambda)$ and $\nu_{05}(\lambda)$ terms. Its amplitude is of the same order of magnitude as that of the $\nu_{05}(\lambda)$ frequency, yet significantly smaller than the dominant contribution from $\nu_{02}(\lambda)$. As a result, the oscillatory behavior of $n(0,t)/n(0,0)$ (see Fig.~\ref{fig:gamma2_N4}(b)) closely follows that of $\gamma^{(2)}(t)/\gamma^{(2)}(0)$, with $\nu_{02}(\lambda)$ being the leading frequency in both cases.

We now turn to the analysis of the Tan contact $\mathcal{C}_T(t)$, which, unlike $n(0,t)$, is diagonal in the common eigenbasis of $\hat{H}_\lambda$ and $\hat{\Gamma}^{(2)}$ (see Eq.~\eqref{eq:CT_decomp}). However, since $c_0^{\rm SU}$, $c_2^{\rm SU}$, and $c_5^{\rm SU}$ are all nonzero, the expression of $\mathcal{C}_T(t)$ becomes more intricate than that of $\gamma^{(2)}(t)$, as it depends on the three populations $|w_0(t)|^2$, $|w_2(t)|^2$, and $|w_5(t)|^2$. Using the relations $|w_0(t)|^2 + |w_2(t)|^2 + |w_5(t)|^2 = 1$ and $|w_0(t)|^2 = \gamma^{(2)}(t)/\gamma_0$, we can rewrite the Tan contact for the $2+2$ bosonic system as:
\begin{equation}
\begin{split}
    \mathcal{C}_T(t) &= c_0^\mathrm{SU} |w_0(t)|^2 + c_2^\mathrm{SU} |w_2(t)|^2 + c_5^\mathrm{SU} |w_5(t)|^2 \\
    &= \mathcal{C}_0 + \left( c^\mathrm{SU}_0 - \frac{1}{2}(c^\mathrm{SU}_2 + c^\mathrm{SU}_5) \right) \frac{\gamma^{(2)}(t)}{\gamma_0} \\
    &\quad + \frac{1}{2}(c^\mathrm{SU}_2 - c^\mathrm{SU}_5)(|w_2(t)|^2 - |w_5(t)|^2),
\end{split}
\label{eq:CT0_N4}
\end{equation}
where $\mathcal{C}_0 = \frac{1}{2}(c^\mathrm{SU}_2 + c^\mathrm{SU}_5)$ is a constant term. The second term is proportional to the symmetry witness $\gamma^{(2)}(t)$, while the third term accounts for the population imbalance between $|w_2(t)|^2$ and $|w_5(t)|^2$, which vanishes at $t=0$. This last contribution arises at second order in $\lambda$ (see Eqs.~\eqref{eq:w2w2} and \eqref{eq:w5w5} in Appendix~\ref{app}), but in this case carries a smaller weight compared to the $\gamma^{(2)}(t)$ term, as it is modulated by the difference $c^\mathrm{SU}_2 - c^\mathrm{SU}_5$, which is proportional to
the energy gap between two states within the same symmetry sector.

The relative variation of the Tan contact then reads:
\begin{equation}
\begin{split}
\dfrac{ \mathcal{C}_T(t) }{ \mathcal{C}_T(0) } &= \dfrac{1}{2} + \dfrac{1}{2} \frac{\gamma^{(2)}(t)}{\gamma_0} + \dfrac{\sqrt{3}}{6}(|w_2(t)|^2 - |w_5(t)|^2) \\
&\simeq 1 + (-0.44 + 0.35 \cos(\nu_{02}(\lambda)t) \\&+ 0.09 \cos(\nu_{05}(\lambda)t)) \lambda^2 + o(\lambda^3),
\end{split}
\label{eq:CT_N4}
\end{equation}
as shown in Fig.~\ref{fig:gamma2_N4}(c), where it is compared with the exact numerical results.

As with $\gamma^{(2)}(t)$ [see Eq.~\eqref{eq:gamma2_N4}], only the $\nu_{02}(\lambda)$ and $\nu_{05}(\lambda)$ frequencies contribute to the Tan contact to second order in perturbation theory. These two terms are of the same order, with the $\nu_{02}(\lambda)$ component being approximately three times larger in amplitude. This makes the behavior of $\mathcal{C}_T(t)/\mathcal{C}_T(0)$ slightly more distinct from that of $n(0,t)/n(0,0)$ when compared to the symmetry witness $\gamma^{(2)}(t)/\gamma^{(2)}(0)$. Nonetheless, the dominant oscillatory features remain qualitatively consistent across all three quantities.

\subsection{Large-$\lambda$ regime}\label{subsec:largel}
We now focus on the opposite case, $\lambda\gg 1$, where the term $\lambda \hat V_{\rm SB}$ dominates with respect to $\hat H_{\rm SU}$.
In this limit, the energy eigenvalues can be read directly from Eq.~(\ref{eq:VSB_diag}), because the off-diagonal elements given by Eq.~\eqref{eq:Hsu_offd} can be neglected, and each snipped state, $\ket{P}$, and its spin-flipped one, $\ket{U(P)}$, correspond to the same eigenvalue. As discussed in Sec.~\ref{subs:sym}, the eigenvectors $\ket{\chi_n^{(\lambda)}}$ are also eigenstates of the spin-flip operator, therefore, we can write $\ket{\chi_n^{(\lambda\gg 1)}} \simeq \ket{\chi^\pm_P} = \frac{1}{\sqrt{2}}(|P\rangle \pm|U(P)\rangle)$ with $+ (-)$ corresponding to even (odd) symmetry under spin flip.  Thus, the initial state, which is the most symmetric, will be given by $|\chi(0)\rangle=\sum_{P=1}^{M/2}(1/\sqrt{M/2})|\chi_{P}^{+}\rangle$, with $M$ the total number of snippets (see Sec.~\ref{sec:model}).

The time evolution of the modulus square of its projection on $|\chi_0^{\rm  SU}\rangle$ takes the form
\begin{equation}
    |w_0(t)|^2= |\langle \chi_0^{\rm  SU}|\chi(t)\rangle|^2=\frac{4}{M^2}\left|\sum_{P=1}^{M/2}e^{-i\omega_{P}t}\right|^2,
    \label{eq:diffr1}
\end{equation}
where the frequencies $\omega_{P}=\lambda[V_{\rm SB}]_{PP}/\hbar=\eta_P\theta$ are integer multiples of $\theta=-2\lambda\alpha/g_{\uparrow\downarrow}$, with $\eta_P=N_{\sigma\sigma}^{(P)}=0,1,\dots N-2$. We then replace the sum over $P$ with the sum over $\eta_P$ in Eq.~\eqref{eq:diffr1} and we omit the subscript $P$ to shorten the notation, such that
\begin{equation}
    |w_0(t)|^2=\frac{4}{M^2} 
\left|\sum_{\eta=0}^{N-2}p_\eta e^{-i\eta\theta t}\right| ^2,
    \label{eq:diffr2}
\end{equation}
where $p_\eta$ is the multiplicity of $\eta $, i.e.,  the number of neighbouring equal-spin pairs of a given snippet. If all the $p_\eta$ are the same,   Eq.~(\ref{eq:diffr2})  is equivalent to that of the light intensity diffracted by a diffraction grating with $N-1$ slits~\cite{Born-and-Wolf}, for which destructive interference leads to black fringes,  which would correspond to zeros of $|\langle \chi_0^{\rm  SU}|\chi(t)\rangle|^2$. For our system, this happens for $N=4$, where $p_\eta=1$ for all $\eta=0,1,2$, and Eq.~(\ref{eq:diffr2}) is zero for $\theta t=\pm \frac{2}{3}\pi+2m\pi$ with $m$ a relative integer (as expected for the 3 slits problem).

For a generic $N$, we can calculate the multiplicity $p_\eta$ using combinatorial arguments. A snippet configuration $(P)$ can be viewed as a sequence of $b= N_{\sigma\bar\sigma}^{(P)}+1$ spin blocks, where $N_{\sigma\bar \sigma}^{(P)}$ is the number of neighboring opposite-spin pairs (see Sec.~\ref{sec:model}). These spin blocks are bounded by either a system edge and a spin domain (a notional boundary between opposite spins) or by two spin domains. Each spin block contains at least one spin. If $b$ is even, we can distribute $N/2$ indistinguishable spin-$\up$ particles among $b/2$ blocks, and $N/2$ indistinguishable spin-$\down$ particles among the remaining $b/2$ blocks. On the other hand, if $b$ is odd, then $N/2$ indistinguishable spin-$\up$ (or spin-$\down$) particles must be placed in $(b+1)/2$ blocks, and $N/2$ indistinguishable spin-$\down$ (or spin-$\up$) particles in the remaining $(b-1)/2$ blocks. We can then treat the two spin components separately, and the multiplicity is obtained by multiplying the number of possible configurations for each component.

To count the number of ways to distribute $N/2$ identical spins into $b/2$ blocks, we observe that there are $N/2 - 1$ possible gaps between adjacent identical spins, in which we can place $b/2 - 1$ dividers (bars) to define the blocks. This gives a total of $\binom{N/2 -1}{b/2 -1}$ configurations, according to the well-known stars-and-bars method in combinatorics~\cite{Feller}. Therefore, if $b$ is even, the total multiplicity is given by the product $\binom{N/2 -1}{b/2 -1}\binom{N/2 -1}{b/2 -1}$. In the case where $b$ is odd, the same reasoning applies, and the multiplicity becomes $\binom{N/2 -1}{(b+1)/2 -1}\binom{N/2 -1}{(b-1)/2 -1}$.

Finally, by rewriting $b$ as $N-\eta$ (see discussion below Eq.~\eqref{eq:Hlambda_diag} in  Sec.~\ref{sec:model} , we find that
\begin{equation}
    p_\eta = 
    \begin{cases}
          \displaystyle\binom{N/2-1}{(N-\eta-2)/2 }^2 & {\rm if} \,\eta \,{\rm even} \\
     \displaystyle\binom{N/2-1}{(N-\eta-3)/2}\binom{N/2-1}{(N-\eta-1)/2} & {\rm if} \,\eta\, {\rm odd}
    \end{cases},
    \label{eq:pk}
\end{equation}
and we have therefore all the parameters needed to calculate Eq.~\eqref{eq:diffr2} for a generic $N$.
\begin{figure}
    \centering
    \includegraphics[scale=0.55]{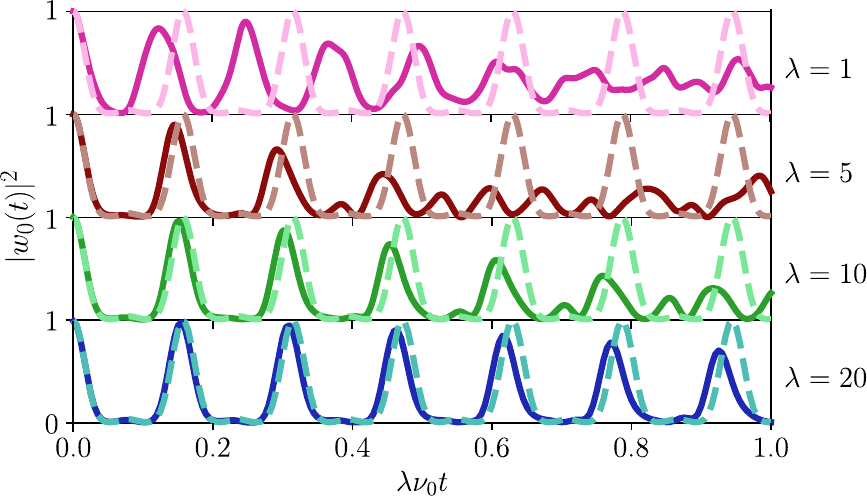}
\caption{Time evolution of the modulus square of the amplitude $w_{0}(t) = \langle \chi_0^{\rm  SU}|\chi(t)\rangle$, namely, the symmetry population for the most symmetric state, ${\cal W}_\gamma(t)$ for the symmetry $\gamma_0$, 
for $N=6$ and different $\lambda = 1, 5, 10, 20$ from top to bottom, as indicated on the right side of the figure.  The dashed lines are the analytical expression presented in Eq.~\eqref{eq:diffr2}. The time axis has been rescaled by $\lambda$. }
\label{fig:biglambda_N6}
\end{figure}

In Fig.~\ref{fig:biglambda_N6} we show the behaviour of $|w_0(t)|^2=|\langle \chi_0^{\rm  SU}|\chi(t)\rangle|^2$ for $N=6$ and increasing $\lambda=1, 5, 10, 20$ compared to the approximated analytical expression given in Eq.~(\ref{eq:diffr2}). We see that by increasing $\lambda$ the exact results approaches the analytical behavior for longer timescales. We also observe that in between to consecutive maxima (that occur when $\theta t=2\pi$ mod $2\pi$)
$|w_0|^2$ is almost zero.
We thus evaluate an upper bound to the minimum of $|w_0(t)|^2$ by calculating its value at $\theta t=\pi$ (mod $2\pi$)
\begin{equation}
\begin{split}
  &  |w_0(t)|^2_{\rm min}\leq \dfrac{4}{M^2}\left|\sum_{\eta=0,1,\dots}^{N-2}p_\eta e^{-i\eta\pi}\right|^2\\
  &= \dfrac{4}{M^2}\left|\sum_{\eta=0,2,\dots}^{N-2}p_\eta-\sum_{\eta=1,3,\dots}^{N-3}p_\eta\right|^2\\
  &= 4\left(\dfrac{(\frac{N}{2})!(\frac{N}{2})!}{N!}\right)^2\left(\dfrac{4^{N/2-1}(\frac{N-3}{2})!}{\sqrt{\pi}(\frac{N}{2})!}\right)^2\\
  &= \dfrac{4^{N-1}}{\pi}\left(\dfrac{(\frac{N}{2})!(\frac{N-3}{2})!}{N!}\right)^2,
    \end{split}
    \label{eq:w0_min}
\end{equation}
that is vanishing for the case of a large number of particles.
Thus we have shown that for $\lambda\gg 1$ and $N\gg 1$, the full symmetric state is fully depleted at certain time $\theta t=(2 m+1)\pi$, with $m$ relative integer.

\subsection{Different initial states}\label{sec:initialcond}
\begin{figure}
    \centering
    \includegraphics[scale=.6]{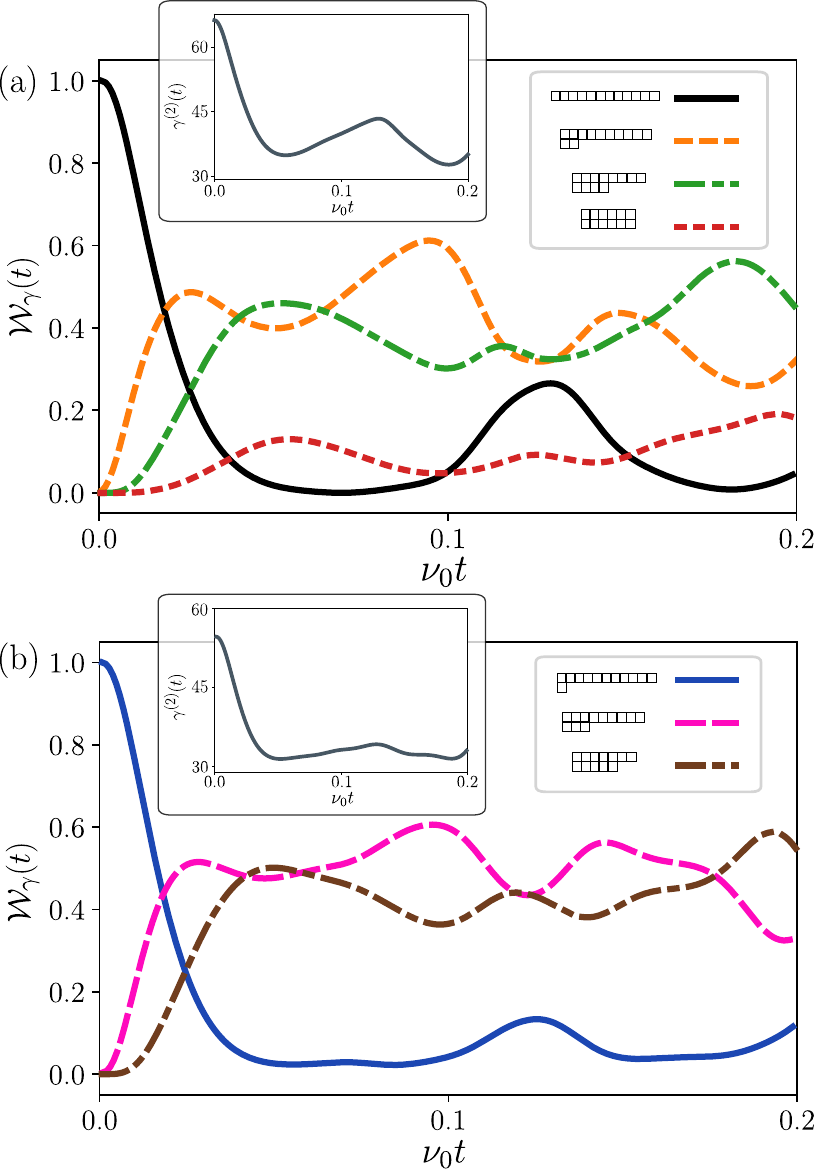}
\caption{
Time evolution of the symmetry populations, ${\cal W}_\gamma(t)$ for $N=12$ bosons with $\lambda = 0.5$.
(a) Initial state $\ket{\chi(0)} = \ket{\gamma_0}$ with $\hat\Gamma^{(2)}$ eigenvalue $\gamma_0 = 66$ corresponding to the Young tableau $(12)$. This state couples to the symmetry sectors $(10,2)$, $(8,4)$, and $(6,6)$, with $\hat\Gamma^{(2)}$ eigenvalues 44, 30, and 24, respectively (see Eq.~\eqref{eq:eigengamma}).
(b) Initial state $\ket{\chi(0)} = \ket{\gamma_1}$ with $\hat\Gamma^{(2)}$ eigenvalue $\gamma_1 = 54$ and Young tableau $(11,1)$. It evolves predominantly through states with symmetries $(9,3)$ and $(7,5)$, whose $\hat\Gamma^{(2)}$ eigenvalues are 36 and 26.
Only nonzero contributions from the involved Young tableau are shown.
Insets: Time evolution of $\gamma^{(2)}(t)$ for each initial state. In (a), $\gamma^{(2)}(t)$ starts at 66 and reaches minima around 35, reflecting the weight of lower-symmetry components. In (b), it starts at 54 and drops to a minimum near 30.
}
\label{fig:diff_initials}
\end{figure}

Finally, in this section, we comment on different choices of initial states. 
As discussed in Sec.~\ref{subs:sym}, the conservation of the spin-flip symmetry allows us to connect only certain types of diagrams in the time evolution of the system.  By choosing as initial state the most symmetric eigenstate of $\hat\Gamma^{(2)}$, only the Young tableaux with the same symmetry under spin flip can be populated, as shown in Sec.~\ref{sec:res} for $N=6$ and in Fig.~\ref{fig:diff_initials}(a) for $N=12$. However, one could make a different choice of initial state, as in Fig.~\ref{fig:diff_initials}(b) for $N=12$  and populate the  other Young tableaux. To start from a different initial state in our calculations is straightforward, however, experimentally, it might be more challenging to start with a well-defined symmetry which is not the most symmetric one that is the ground state of $\hat{H}_{SU}$.  

\section{Conclusions}\label{sec:conclu}
In this work, we have extended the analysis of symmetry oscillations—first introduced in Ref.~\cite{art:musolino2024} to a broader class of models and interaction regimes. Our study confirms that this phenomenon is not restricted to a specific model or fine-tuned parameters, but instead constitutes a robust and universal dynamical feature of multicomponent strongly interacting 1D quantum mixtures.

We have identified the selection rules governing the coupled symmetry sectors during the dynamics: in the case of balanced mixtures, the Hamiltonian preserves spin-flip symmetry, which constrains the symmetry oscillations to occur between states with the same parity under spin inversion. This insight provides a deeper understanding of the structure of the oscillatory dynamics.

We have also characterized how the amplitude and visibility of symmetry oscillations depend on the magnitude of the symmetry-breaking perturbation. In the regime of weak symmetry breaking, a second-order perturbative analysis captures the essential features of the dynamics and provides clear guidance for experimental realizations in systems where tuning the inter- and intraspecies interactions is limited~\cite{arX:Beugnon2024, arX:Zhao2025}.

In the opposite regime of strong symmetry breaking, we have shown that the population of the initial symmetry sector can be completely depleted—leading to a maximal contrast in the oscillations. This effect persists in the thermodynamic limit and can be interpreted as a form of destructive interference, reminiscent of an $N{-}1$-slit diffraction pattern in symmetry space.

Finally, this mechanism could be exploited to engineer quantum states with nontrivial symmetry properties: by halting the dynamics at a specific time (e.g., via a sudden interaction quench), one can selectively prepare many-body states belonging to exotic irreducible representations of the symmetric group. This opens new avenues for the controlled generation of unconventional quantum states with tailored permutation symmetries.

\begin{acknowledgments}
All the authors warmly thank Fr\'ed\'eric H\'ebert for
the derivation of Eq.~\eqref{eq:pk} and  Jean Dalibard for fruitful  discussions.  We acknowledge  financial support from the ANR-21-CE47-0009 Quantum-SOPHA project.  M.A., P.V., and A.M. also acknowledge the financial support from  the
ANR-23-PETQ-0001 Dyn1D at the title of France 2030.
\end{acknowledgments}

\appendix
\section{Explicit expressions for the case $N=4$}
\label{app}
In this Appendix, we provide additional explicit expressions  for the results presented in Sec.~\ref{sec:perturbative}.
\subsection{Eigenvectors and eigenvalues}
The eigenvectors of the SU$(2)$ Hamiltonian  for the case of a 2+2 bosons trapped in a box trap (Eq.~\eqref{eq:Hz_pert} with $\lambda = 0$) are given in the snippet basis by 
\begin{widetext}
\begin{equation}
    \begin{split}
        |\chi_0^{\rm SU}\rangle &= \frac{1}{\sqrt{6}} (1, 1, 1, 1, 1, 1),\\\allowdisplaybreaks
        |\chi_1^{\rm SU}\rangle &= \Big( -\sqrt[4]{\frac{2}{11}}, -\sqrt{\frac{1}{2}-\sqrt{\frac{2}{11}}}, 0, 0, \sqrt{\frac{1}{2}-\sqrt{\frac{2}{11}}}, \sqrt[4]{\frac{2}{11}}\Big),\\\allowdisplaybreaks
        |\chi_2^{\rm SU}\rangle &= \Big( \frac{1}{2} \left(\frac{1}{\sqrt{2}} + \frac{1}{\sqrt{6}}\right),\frac{1}{2} \left(-\frac{1}{\sqrt{2}} + \frac{1}{\sqrt{6}}\right),-\frac{1}{\sqrt{6}},-\frac{1}{\sqrt{6}}, \frac{1}{2} \left(-\frac{1}{\sqrt{2}} + \frac{1}{\sqrt{6}}\right),\frac{1}{2} \left(\frac{1}{\sqrt{2}} + \frac{1}{\sqrt{6}}\right)\Big),\\\allowdisplaybreaks
        |\chi_3^{\rm SU}\rangle &= 
    \frac{1}{\sqrt{2}}(0, 0, -1, 1, 0, 0),\\\allowdisplaybreaks
    |\chi_4^{\rm SU}\rangle &=\Big( -\sqrt{\frac{1}{2}-\sqrt{\frac{2}{11}}}, \sqrt[4]{\frac{2}{11}}, 0, 0, -\sqrt[4]{\frac{2}{11}}, \sqrt{\frac{1}{2}-\sqrt{\frac{2}{11}}}\Big),\\\allowdisplaybreaks
    |\chi_5^{\rm SU}\rangle &=\Big(\frac{1}{2} \left(\frac{1}{\sqrt{2}} - \frac{1}{\sqrt{6}}\right),-\frac{1}{2} \left(\frac{1}{\sqrt{2}} + \frac{1}{\sqrt{6}}\right),\frac{1}{\sqrt{6}},\frac{1}{\sqrt{6}}, -\frac{1}{2} \left(\frac{1}{\sqrt{2}}+ \frac{1}{\sqrt{6}}\right),\frac{1}{2} \left(\frac{1}{\sqrt{2}} - \frac{1}{\sqrt{6}}\right)\Big).
    \end{split}
    \label{eq:eig_SU}
\end{equation}
\end{widetext}
To second order in perturbation theory,  using Eq.~\eqref{eq:chinlam}, the normalized eigenvectors of $H_\lambda$ for $N=4$ (Eq.~\eqref{eq:Hz_pert}) can be written in terms of the SU-eigenvectors (Eq.~\eqref{eq:eig_SU}) as
\begin{widetext}
\begin{equation}
\begin{split}
\ket{\chi_0^{(\lambda)}} &=  (1-0.44 \lambda^2)\ket{\chi_0^\mathrm{SU}} +(0.91 \lambda + 0.83 \lambda^2)\ket{\chi_2^\mathrm{SU}} + (0.24\lambda -0.06 \lambda^2) \ket{\chi_5^\mathrm{SU}} + o(\lambda^3), \\\allowdisplaybreaks
\ket{\chi_1^{(\lambda)}} &=(1-0.12 \lambda^2)\ket{\chi_1^\mathrm{SU}}+(0.5\lambda-0.5 \lambda^2 )\ket{\chi_4^\mathrm{SU}}+ o(\lambda^3),\\\allowdisplaybreaks
\ket{\chi_2^{(\lambda)}}&= (1-0.41\lambda^2)\ket{\chi_2^\mathrm{SU}} - (0.91\lambda + 0.83\lambda^2 ) \ket{\chi_0^\mathrm{SU}}  - 0.30 \lambda^2\ket{\chi_5^\mathrm{SU}}+ o(\lambda^3),\\
\ket{\chi_3^{(\lambda)}} &= \ket{\chi_3^\mathrm{SU}} + o(\lambda^3),\\\allowdisplaybreaks
\ket{\chi_4^{(\lambda)}} &= (1- 0.12\lambda^2)\ket{\chi_4^\mathrm{SU}} + (-0.5\lambda +0.5 \lambda^2 )\ket{\chi_1^\mathrm{SU}}+ o(\lambda^3),\\\allowdisplaybreaks
\ket{\chi_5^{(\lambda)}} &= (1-0.03 \lambda^2)\ket{\chi_5^\mathrm{SU}} +(- 0.24\lambda +0.06 \lambda^2 )\ket{\chi_0^\mathrm{SU}} +0.08 \lambda^2\ket{\chi_2^\mathrm{SU}} + o(\lambda^3),
 \label{eq:eigenvec_2nd_norm}
\end{split}
\end{equation}
\end{widetext}
with corresponding eigenvalues  in units of $g_{\up \down}/\alpha$
\begin{equation}
\begin{split}
\epsilon_0(\lambda) &= -( 6 + 2\lambda + \frac{4}{3} \lambda^2 + o(\lambda^3)), \\\allowdisplaybreaks
\epsilon_1(\lambda) &= -(4+\sqrt{2} + 3.41 \lambda + \frac{1}{\sqrt{2}} \lambda^2 + o(\lambda^3)),\\\allowdisplaybreaks
\epsilon_2(\lambda) &= -(3+\sqrt{3} + 3.15\lambda -  1.05 \lambda^2 + o(\lambda^3)),\\\allowdisplaybreaks
\epsilon_3(\lambda) &= -(4 + 2\lambda  + o(\lambda^3)),\\\allowdisplaybreaks
\epsilon_4(\lambda) &=-( 4-\sqrt{2} + 0.58\lambda -  \frac{1}{\sqrt{2}} \lambda^2 + o(\lambda^3)),\\\allowdisplaybreaks
\epsilon_5(\lambda) &= -(3-\sqrt{3} + 0.84\lambda - 0.28 \lambda^2 + o(\lambda^3)).
\end{split}
\label{eq:eigenvalues_N4}
\end{equation}
\subsection{Second-order perturbative expansion of the eigenstate populations}
Here we provide the explicit expressions of the second-order perturbative expansion of the terms $|w_0(t)|^2$, $|w_2(t)|^2$, $|w_5(t)|^2$ and $2{\rm Re}[w_2^*(t)w_5(t)]$ entering in Eqs.~\eqref{eq:gamma2_pert}, \eqref{eq:CT0_N4} and \eqref{eq:n0_N4b}, as obtained by using Eqs.~\eqref{eq:wei_pert_N4}:
\begin{equation}
\begin{split}
    |w_0(t)|^2&=1 + (-1.76 + 1.656 \cos(\nu_{02}(\lambda)t) \\&+ 0.115 \cos(\nu_{05}(\lambda)t))\lambda^2+o(\lambda^3),
    \label{eq:w0w0}
    \end{split}
\end{equation}
\begin{equation}
    |w_2(t)|^2= 1.65(1 - \cos(\nu_{02}(\lambda)t))\lambda^2+o(\lambda^3),
      \label{eq:w2w2}
\end{equation}
\begin{equation}
    |w_5(t)|^2= 0.12(1 - \cos(\nu_{05}(\lambda)t))\lambda^2+o(\lambda^3),
      \label{eq:w5w5}
\end{equation}
\begin{equation}
\begin{split}
    2{\rm Re}[w_2^*(t)w_5(t)]&=0.44(1 - \cos(\nu_{02}(\lambda)t) - \cos(\nu_{05}(\lambda)t)\\& + \cos(\nu_{25}(\lambda)t))\lambda^2+o(\lambda^3).
      \label{eq:w2w5}
    \end{split}
\end{equation}

\bibliographystyle{apsrev4-2}
%\bibliography{bibliog}

%

\end{document}